\documentclass[letterpaper]{sae}
\usepackage{color}
\usepackage{fancyhdr}
\usepackage{lastpage}
\usepackage{etoolbox}

\PaperTitle{Experimental Validation of Eco-Driving and Eco-Heating Strategies for Connected and Automated HEVs} 
\AddAuthor{M. Reza Amini$^1$, Qiuhao Hu$^1$, Hao Wang$^2$, Yiheng Feng$^3$, Ilya Kolmanovsky$^1$, and Jing Sun$^1$}{\vspace{2pt}$^1$University of Michigan, Ann Arbor, MI, USA\\$^2$Ford Motor Company, Dearborn, MI, USA\\$^3$Purdue University, West Lafayette, IN, USA}
\PaperNumber{2021-01-0435}
\SAECopyright{2021}

\usepackage[utf8]{inputenc}
\usepackage[english]{babel}
\usepackage{graphicx}
\usepackage[utf8]{inputenc}
\usepackage{algorithm}
\usepackage[noend]{algpseudocode}
\usepackage{amsmath}
\usepackage{array}
\usepackage{booktabs}
\usepackage{nomencl}
\usepackage{rotating}
\usepackage{indentfirst}
\usepackage{amsmath}
\usepackage{amssymb}
\usepackage{hyperref}
\usepackage{setspace}
\usepackage{extdash}
\usepackage{graphicx}
\usepackage{graphics}
\usepackage{epstopdf}
\usepackage{lipsum}
\usepackage{multirow}
\usepackage{tablefootnote}
\usepackage[usenames,dvipsnames]{xcolor}
\usepackage{cite}
\usepackage{mathtools}
\usepackage{mathrsfs}
\usepackage{makecell}

\begin{document}
\graphicspath{{figures/}}

\pagestyle{fancy}
\fancyhf{}
\renewcommand{\headrulewidth}{0pt}
\fancypagestyle{firstpage}{\lhead{\vspace{-0.75 cm} \small \textbf{{\fontfamily{cmss}\selectfont
Citation: \textcolor{blue}{Amini, M.R., Hu, Q., Wang, H., Feng, Y., Kolmanovsky, I., Sun, J., ``{Experimental Validation of Eco-Driving and Eco-Heating Strategies for Connected and Automated HEVs},'' SAE Technical Paper 2021-01-0435, 2021.\\ \url{{https://saemobilus.sae.org/content/2021-01-0435}}}}}}}
\fancyfoot[L]{Page \thepage \hspace{1pt} of \pageref{LastPage}} 

\maketitle 

\thispagestyle{firstpage}

\AtEndEnvironment{algorithm}{\kern2pt\hrule\relax\vskip3pt\@algcomment}

\newcommand\algcomment[1]{\def\@algcomment{\footnotesize#1}}

\pagestyle{fancy}
\fancyhf{}
\renewcommand{\headrulewidth}{0pt} 
\fancyfoot[L]{Page \thepage \hspace{1pt} of \pageref{LastPage}} 

\section{Abstract} \vspace{-8pt}
This paper presents experimental results that validate eco-driving and eco-heating strategies developed for connected and automated vehicles (CAVs). By exploiting vehicle-to-infrastructure (V2I) communications, traffic signal timing, and queue length estimations, optimized and smoothed speed profiles for the ego-vehicle are generated to reduce energy consumption.~
%
Next, the planned eco-trajectories are incorporated into a real-time predictive optimization framework that coordinates the cabin thermal load (in cold weather) with the speed preview, i.e., eco-heating. To enable eco-heating, the engine coolant (as the only heat source for cabin heating) and the cabin air are leveraged as two thermal energy storages. Our eco-heating strategy stores thermal energy in the engine coolant and cabin air while the vehicle is driving at high speeds, and releases the stored energy slowly during the vehicle stops for cabin heating without forcing the engine to idle to provide the heating source. To test and validate these solutions, a power-split hybrid electric vehicle (HEV) has been instrumented for cabin thermal management, allowing to regulate heating, ventilation, and air conditioning (HVAC) system inputs (cabin temperature setpoint and blower flow rate) in real-time. Experiments were conducted to demonstrate the energy-saving benefits of eco-driving and eco-heating strategies over real-world city driving cycles at different cold ambient temperatures. The data confirmed average fuel savings of 14.5\% and 4.7\% achieved by eco-driving and eco-heating, respectively, offering a combined energy saving of more than 19\% when comparing to the baseline vehicle driven by a human driver with a constant-heating strategy.


%
\vspace{-4pt}
\section{Introduction}\label{sec:sec1}\vspace{-8pt}
Connectivity and automated driving are emerging technologies that are transforming the transportation system landscape. They have opened up unprecedented opportunities for enhanced safety, mobility, and efficiency~\cite{rios2016survey,guanetti2018control,GM_nextcar} through data-sharing, active perception and localization, and onboard sensing, computation, planning, and control. In particular, CAVs can make real-time decisions by exploiting increased situational awareness gained from sensory data. While the main focus in the development of CAVs has been on driving safety and comfort, CAVs implications on energy consumption have also been considered, see~\cite{taiebat2018review,gawron2018life,vahidi2018energy,wei2020review} and the references therein. Most of the studies thus far have focused on reducing the traction power losses (e.g., eco-driving, platooning, cooperative adaptive cruise control). The optimization of thermal loads (including those of engine, cabin, and battery), on the other hand, has received less attention.

\vspace{-4pt}
The efficient thermal management of vehicles can have a significant impact on energy consumption, powertrain efficiency, and passenger comfort~\cite{lajunen2020review,jeffers2015climate}. The impact of thermal management is even more pronounced for electrified vehicles with larger thermal loads due to the higher power density of the electrified systems onboard~\cite{alleyne2018power}.~For CAVs, the situational awareness and look-ahead previews gained from vehicle-to-infrastructure (V2I) and vehicle-to-vehicle (V2V) communication can be exploited for coordination of power and thermal loads. As shown in previous studies~\cite{Amini_TCST_2019,AminiACC19,Amini_TCST_2020,BruceACC20,SAE_2020}, integrated power and thermal management (iPTM) of CAVs can greatly benefit from leveraging the coupling between power and thermal loads and accounting for the timescale separation between power and thermal dynamic responses. Along these lines, energy-efficient strategies for cooling (i.e., eco-cooling) of cabin~\cite{hwang2018,Amini_TCST_2019,cvok2020optimization,schaut2019thermal} and battery~\cite{Amini_TCST_2020,AminiCDC18,amini2020long,masoudi2017mpc}, as well as iPTM strategies for co-optimization of engine, cabin, and aftertreatment systems~\cite{SAE_2020,BruceACC20,Amini_CCTA19,gong2019integrated,AminiACC19} have been developed. When conflated with technologies focused on traction power optimization (e.g., eco-driving), efficient thermal management of CAVs is shown to have the potential for delivering fuel-savings of up to 18-20\%~\cite{Amini_CCTA19,AminiACC19,Atkinson_arpae}. 

\vspace{-4pt}
While previous publications have been touting the energy-saving potential of iPTM, many if not all of the previous approaches are demonstrated using simulations.  In contrast, in this paper,~we not only propose but also experimentally validate an energy-efficient eco-heating strategy for cabin thermal management of hybrid electric connected and automated vehicles (HECAVs) operating at cold ambient temperatures. Further, we present eco-driving results for congested city driving scenarios. We use real-world drive cycles and experimental testing to confirm the repeatability of the~cumulative fuel-savings achieved through our eco-driving and eco-heating strategies for CAVs.

\begin{figure}[t!]
\begin{center}
\includegraphics[angle=0,width= \columnwidth]{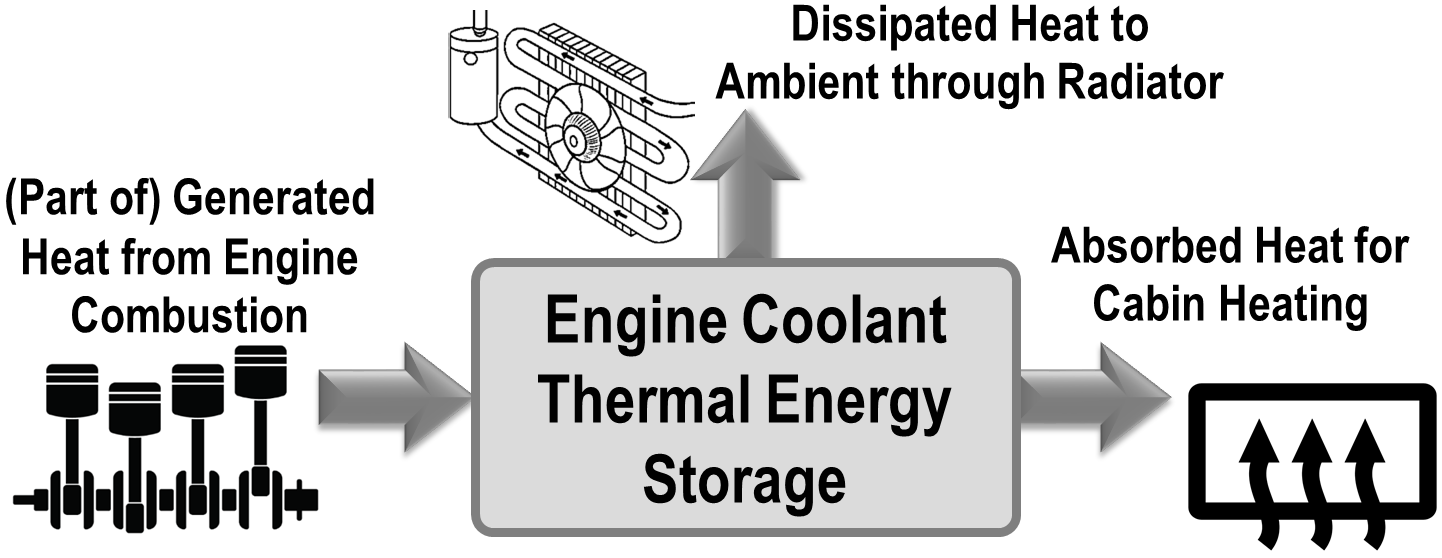} \vspace{-0.75cm}
\textcolor{Blue}{\caption{\label{fig:ThermalStorage} The energy balance of the engine coolant treated as thermal energy storage.}}\vspace{-4pt}
\end{center}
\end{figure}

\vspace{-4pt}
For hybrid electric vehicles (HEVs) operating at cold ambient, thermal management of the engine and cabin heating are the main thermal loads. For HEVs considered in this paper, the combustion engine is the only source for cabin heating. A portion of the generated heat during the combustion process is absorbed by the engine coolant while the rest of the fuel energy is converted to either mechanical work or wasted through the exhaust gases. The stored thermal energy in the coolant is then used to heat the cabin air at the heater cores if there is a demand for cabin heating. Figure~\ref{fig:ThermalStorage} shows the energy balance for the engine coolant, with the heat generated by the engine being the only input. As can be seen, the absorbed heat is either dissipated to the ambient and cabin via the radiator and heater cores, respectively, or stored within the coolant as thermal energy, resulting in the coolant temperature rise. Thus, the coolant serves as thermal energy storage and its stored thermal energy can be controlled by regulating the input and outputs.

\vspace{-4pt}
Coolant temperature is conventionally controlled passively through a thermostat that directs the coolant through the radiator path whenever its temperature reaches a specific threshold (e.g., 80$^oC$). Cabin heating, on the other hand, depletes thermal energy from the coolant at the heater cores and reduces the coolant temperature. Once the coolant temperature drops to below a certain threshold (e.g.~40-50$^oC$~\cite{gong2019integrated,SAE_2020}), the engine is commanded to run to generate heat---even if there is no demand for driving. Such thermostat-like regulation of the coolant temperature leads to inefficient use of the coolant as thermal energy storage. Note that storing and releasing thermal energy in coolant depend on its input (supply) and outputs (demand), similar to a battery that stores electric energy. If the supply and demand sides of the thermal energy storage are coordinated optimally, the wasted heat through radiator and engine idling can be minimized to save fuel. Look-ahead information gained from V2I/V2V (V2X) communications can facilitate such optimal coordination for thermal energy storage. Multiple competing objectives (e.g., fuel-saving versus cabin heating), constraints (e.g., the lower and upper limits for coolant temperature), and the need to integrate the look-ahead information in this problem can be addressed using Model Predictive Control (MPC)~\cite{rawlings2017model}.

\vspace{-4pt}
The look-ahead information used by our eco-heating strategy is obtained from V2I data. To this end, we use the eco-trajectory planning algorithm developed in~\cite{yang2019eco} to generate vehicle speed trajectories in real-time to reduce energy consumption. The approach in~\cite{yang2019eco} accounts for vehicle queuing dynamics at signalized intersections and results in eco-arrival and eco-departure (EAD) behaviors. This generated trajectory is (i) used to conduct eco-driving experiments, and (ii) leveraged as a look-ahead preview by MPC-based eco-heating strategy. Note that, other approaches have also been developed for short- and long-term predictions of vehicle speed using V2X~\cite{amini2020long,prakash2016assessing,Hosseini_PhD,lang2014prediction,hosseini2020prediction,park2014intelligent,hosseini2020use,jiang2016vehicle,ma2015long} which can be integrated with our eco-heating
strategy. 

\begin{figure}[t!]
\begin{center}
\includegraphics[angle=0,width= 0.95\columnwidth]{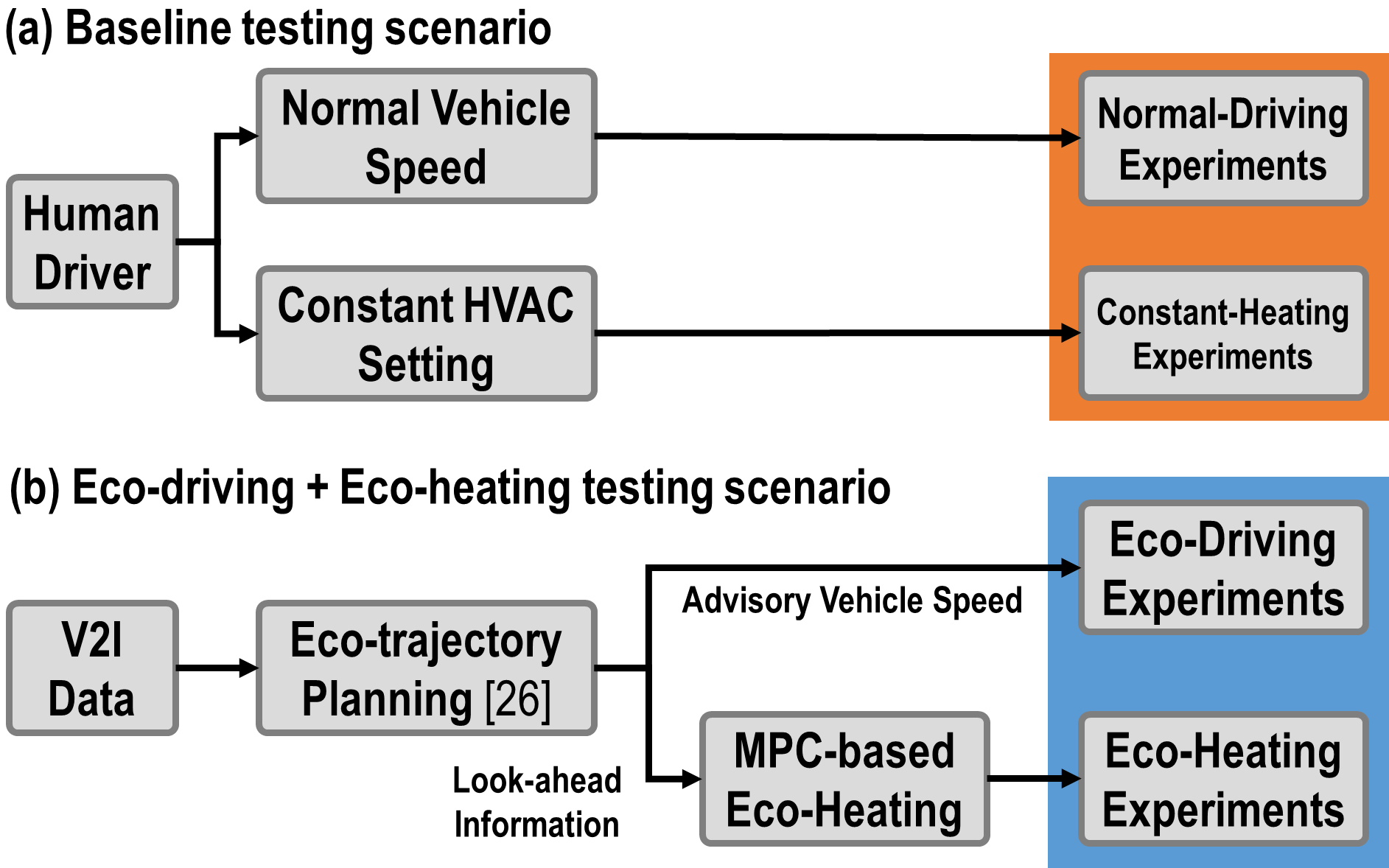} \vspace{-0.35cm}
\textcolor{Blue}{\caption{\label{fig:TestScenario} The testing scenarios presented in this paper.}}\vspace{-4pt}
\end{center}
\end{figure}

\vspace{-4pt}
The testing scenarios for which experimental results are reported in this paper are summarized in Figure.~\ref{fig:TestScenario}. For baseline testing (Figure.~\ref{fig:TestScenario}-(a)), we consider a human driver traveling through an arterial corridor with a normal speed, referred to as normal-driving experiments. The heating, ventilation, and air conditioning (HVAC) system of the vehicle is set at constant set points to provide constant-heating at cold ambient temperature in these normal-driving experiments. The eco-driving and eco-heating strategies, on the other hand, are tested using the vehicle speed profiles generated by the eco-trajectory planning algorithm in~\cite{yang2019eco} and our MPC-based eco-heating approach (Figure.~\ref{fig:TestScenario}-(b)).  

\vspace{-4pt}
The rest of the paper is organized as follows: Firstly, the eco-trajectory planning strategy is briefly reviewed. Secondly, the eco-heating strategy that has been developed is discussed in detail. Next, the experimental setup and test vehicle instrumentation are discussed. The experimental results are presented next, followed by the summary and concluding remarks in the last section. The list of nomenclature and acronyms used in the paper are provided in the Appendix section.

\vspace{-4pt}
\section{Trajectory Planning for Eco-Driving}\vspace{-4pt}
The approach used in this paper for vehicle speed planning is based on our previous works~\cite{yang2019eco,AminiACC19}. By leveraging traffic signal and queue length estimation based on V2I communications, the eco-trajectory planning strategy generates a speed profile with the objective of minimizing energy consumption. A key part of this strategy is the prediction of the queue length using the trajectories of connected vehicles inferred from Basic Safety Messages (BSMs) and from loop-detectors installed on the infrastructure side. The shockwave profile model (SPM)~\cite{wu2011shockwave} is used for modeling the vehicle queuing process. The SPM model provides a green window, which is defined as the time interval during which an ego-vehicle can pass through a given intersection. One particular feature of this eco-trajectory planning algorithm is its ability to predict the queuing dynamics and estimate the green window before the arrival of the ego-vehicle at an intersection. Based on the predicted queue length, signal status, and remaining time, four different eco-driving strategies were designed in~\cite{yang2019eco}, including (i) cruise, (ii) speed up, (iii) slow down, and (iv) stop, see~\cite{yang2019eco} for more details. We use the generated eco-trajectories as:\vspace{-12pt}
\begin{itemize}
    \item advisory speed for eco-driving experiments,\vspace{-6pt}
    \item vehicle speed preview incorporated in MPC for eco-heating experiments. 
\end{itemize}
\begin{figure}[t!]
\begin{center}
\includegraphics[angle=0,width= \columnwidth]{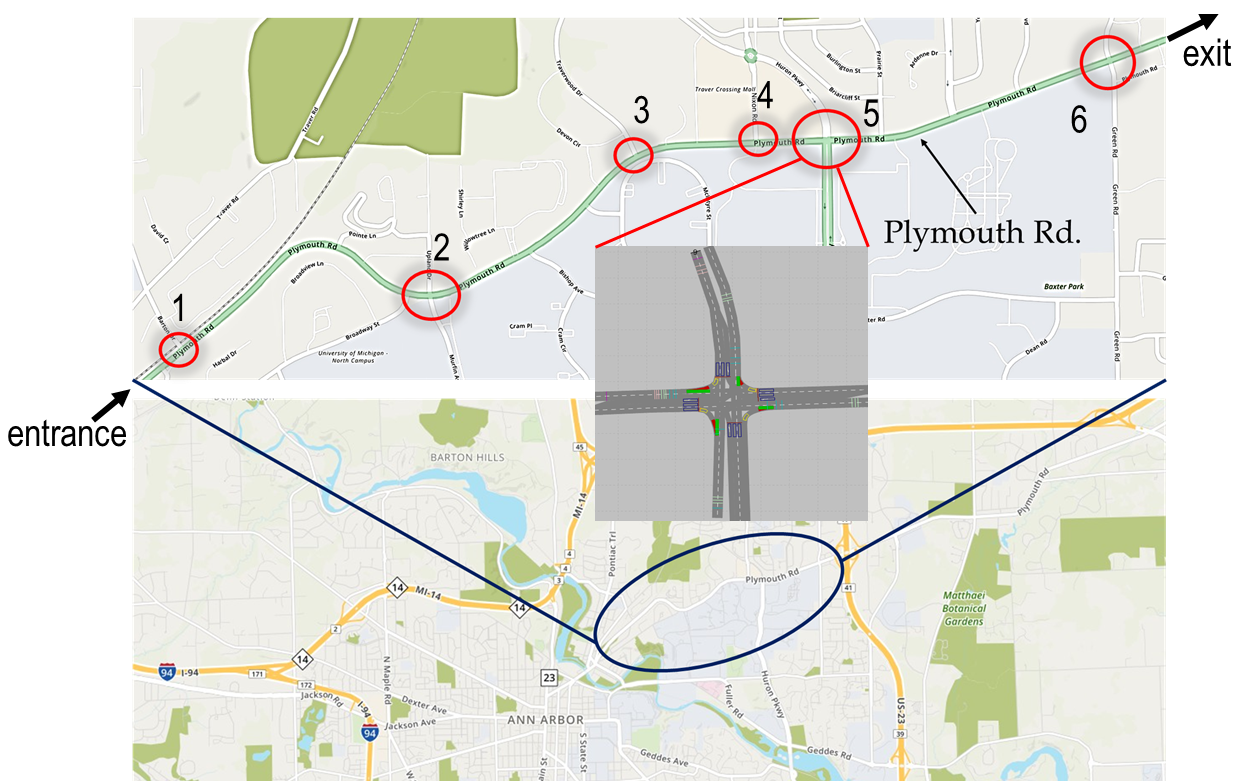} \vspace{-0.75cm}
\textcolor{Blue}{\caption{\label{fig:PlymouthRd} The schematic of Plymouth Rd. in Ann Arbor, MI, with six intersections considered for traffic modeling, eco-trajectory generation, and eco-driving and eco-heating experiments.}}\vspace{-4pt}
\end{center}
\end{figure}

\vspace{-12pt}
{We consider an arterial road corridor with six intersections as the case study. The corridor represents a segment of the Plymouth Rd. in Ann Arbor, MI with a length of 2.2 miles and has two lanes in each direction. This road segment, which is one of the busiest local commuting routes, is shown in Figure~\ref{fig:PlymouthRd} where the red circles denote the location of the intersections, with the entrance to the corridor being on the left (west) side. Real-world data (including traffic volume, turning ratio, and traffic signal timing at each intersection) have been collected during afternoon rush hour (4:00-5:00 PM) to calibrate the VISSIM simulation model~\cite{yang2019eco}, representing a congested traffic condition. The data collected from traffic signals and vehicles are sent to the queuing profile algorithm for green window prediction, which is then used in the eco-trajectory planning algorithm. 

\vspace{-4pt}
The benefits of eco-driving are quantified for HEVs traveling through Plymouth Rd. in our previous works, where a physics-based and experimentally validated model of a power-split HEV was used for energy consumption analysis in MATLAB/Simulink\textsuperscript{\textregistered} environment.~The simulation results in~\cite{AminiACC19,Amini_CCTA19} showed an average fuel saving of 13$\%$ achieved through eco-driving as compared to the baseline case, where the same vehicles were assumed to be driven by human drivers under similar traffic conditions. For more details, see~\cite{yang2019eco,AminiACC19,Amini_CCTA19}.

\vspace{-4pt}
\section{Eco-Heating Strategy}\vspace{-4pt}
The heat supplied by a fully warmed-up engine during the vehicle's driving operation often exceeds the demand for cabin heating.~On the other hand, when the vehicle stops or drives at low speeds, it is preferred to keep the engine off and operate the vehicle using battery only. In the latter case, the coolant temperature decreases rapidly due to the demand for cabin heating while there is no heat supply from the engine, see Figure~\ref{fig:ThermalStorage}. If the coolant temperature drops below a threshold, e.g., 50$^oC$, the engine has to be turned on to provide heat to the coolant even if the vehicle is still at the stop. However, if the coolant temperature is raised when the vehicle comes to stop, the need to idle the engine during the stop can be delayed or avoided as there is more thermal energy stored in the coolant for cabin heating. Alternatively, the demand (HVAC) side can be also manipulated to draw less energy from the coolant during the vehicle stops. The latter strategy may intervene with passengers' comfort if less energy delivered to the cabin leads to lower cabin temperature. To manage this conflict, one can deliver more heat to the cabin (i.e., cabin pre-heating) before the vehicle's stop. Considering the slow thermal dynamics of the cabin air temperature, this pre-heating strategy could avoid or delay engine idling during the stops while maintaining a relatively comfortable cabin thermal environment. The two solutions discussed here are the core ideas used for developing eco-heating strategy. 

\vspace{-4pt}
The eco-heating strategy is algorithmically defined by the solution of the following finite-time MPC problem: \vspace{-4pt}
\begin{equation} 
\begin{aligned}\label{eq:Eco-cooling}
&\min_{\substack{\dot{m}_{bl}\\ T_{s.p.}}} && \sum_{i=0}^{N_p}  \begin{gathered} \Big(P_{DAHP}(i|k)-\beta(i|k)\cdot P_{DAHP}^{targ}(i|k)\Big)^2 \end{gathered},\\
& \text{subject to}
& & T_{cl}(i+1|k)=T_{cl}(i|k),
\end{aligned}
\end{equation}
\begin{equation} 
\begin{aligned}
&
& &\dot{m}_{bl}^{min}\leq \dot{m}_{bl}(i|k)\leq \dot{m}_{bl}^{max},\\
&
& &T_{s.p.}^{max}\leq T_{s.p.}(i|k)\leq T_{s.p.}^{max},\\
&
& & T_{cl}(0|k)=T_{cl}(k).\nonumber
\end{aligned}
\end{equation}
where $\dot{m}_{bl}$ and $T_{s.p.}$, as the optimization variables and inputs to the HVAC system, are the blower air mass flow rate and cabin temperature setpoint, respectively. $\dot{m}_{bl}^{min}$ and $\dot{m}_{bl}^{max}$ are the limits on the HVAC blower flow rate which are governed by the blower motor physical limits. $T_{s.p.}^{min}$, $T_{s.p.}^{max}$ are the lower and upper limits on the cabin temperature setpoint, respectively. The coolant temperature is denoted by $T_{cl}$. Further, $k$ is the index of the discrete sampling instants, $N_p$ is the prediction horizon, during which the look-ahead information is implemented, and $(i|k)$ designates the prediction for the time instant
$k+i\delta t$ made at the time instant $k$ with $\delta t$ being the sampling time. Every time the optimization problem (\ref{eq:Eco-cooling}) is solved, the first element of the computed control inputs are applied to the system, and the prediction horizon is shifted forward by one time-step. The cost function of the MPC in Eq. (\ref{eq:Eco-cooling}) is defined as a quadratic term to track the target \textit{discharge air heating power ($P_{DAHP}^{targ}$)}. The discharge air heating power ($P_{DAHP}$) is a metric defined to quantify cabin heating performance~\cite{Hao_TCST_2020} and facilitate the controller design, which is defined as follows:
\vspace{-4pt}
\begin{eqnarray}
\label{eq:DAHP}
P_{DAHP}(k)=c_p(T_{ain}(k)-T_{amb}(k))\dot{m}_{bl}(k),
\end{eqnarray} 
where $c_p$ is the specific heat capacity of air, $T_{amb}$ is the ambient temperature, $T_{ain}$ is the vent air temperature, and $\dot{m}_{bl}$ is the HVAC blower airflow rate. Note that the $P_{DAHP}$ in Eq. (\ref{eq:DAHP}) is defined for the case when HVAC is running in no-recirculation (fresh air) mode. The integral of $P_{DAHP}$ over time is referred to as the discharge air heating energy ($E_{DAHE}$). We use $E_{DAHE}$ to quantify the overall heating delivered to the cabin over a specified time window. 

\vspace{-4pt}
The target heating power trajectory ($P_{DAHP}^{targ}$) in the MPC cost function (Eq.~(\ref{eq:DAHP})) represents the heating power trajectory delivered by a baseline HVAC controller with constant inputs ($\dot{m}_{bl}$, $T_{s.p.}$). The reason to design the eco-trajectory MPC (\ref{eq:Eco-cooling}) in a tracking scheme to follow $P_{DAHP}^{targ}$ is to ensure a similar cabin heating energy is delivered, as compared to the baseline case with constant HVAC setting. To incorporate the look-ahead information, the variable $\beta$ is added to the cost function of the MPC in Eq.~(\ref{eq:Eco-cooling}). If $\beta=1$, the computed HVAC inputs by the MPC will be constant and the same as the baseline case that delivers $P_{DAHP}^{targ}$. In this paper, however, we define $\beta$ as a function of the vehicle speed~($v$) according to the one-dimensional loop-up table shown in Figure~\ref{fig:beta_finction}. To this end, $\beta$ has been calibrated to deliver the same $E_{DAHE}$ on average as a constant-heating strategy ($\beta=1$) over the Plymouth Rd. corridor shown in Figure~\ref{fig:PlymouthRd}. As can be seen in Figure~\ref{fig:beta_finction}, 
$\beta$ is larger than one for $v\geq20~mph$, allowing for cabin pre-heating. The lowest value of $\beta$ is assigned to vehicle stops. \vspace{-6pt}
\begin{figure}[h!]
\begin{center}
\includegraphics[angle=0,width=1\columnwidth]{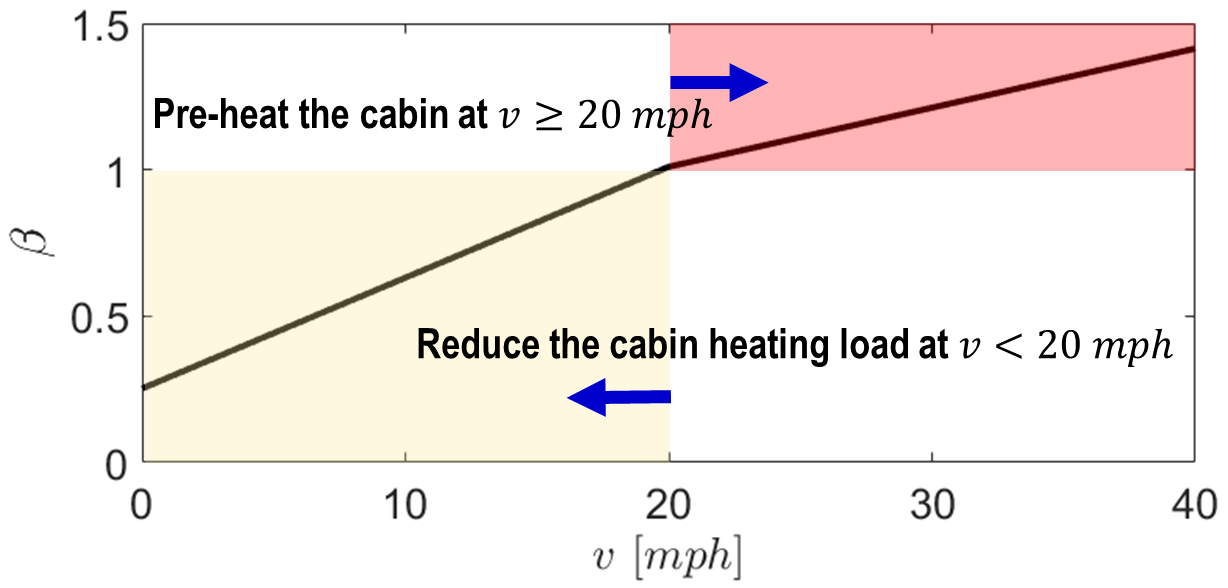} \vspace{-0.85cm}
\textcolor{Blue}{\caption{\label{fig:beta_finction}The look-up table for variable $\beta$ a s function of the vehicle speed.}}\vspace{-7pt}
\end{center}
\end{figure}

\vspace{-4pt}
It should be noted that according to Eq.~(\ref{eq:Eco-cooling}), the coolant temperature is assumed to be constant over the prediction horizon $N_p$.~{Coolant temperature is a function of the HEV power-split strategy that governs the amount of generated heat by the engine. In our experiments, we do not have access to override and control the vehicle power-split strategy, thus predicting the coolant temperature may not be possible when the engine use strategy is unknown over the prediction horizon.}~Its value is updated at each time step based on the real-time measurement from the coolant temperature sensor.

\vspace{-4pt}
\section{Vehicle Testing}\vspace{-4pt}
In this section, first, the test vehicle is described. Next, the eco-driving testing procedure and adoption of the eco-heating strategy for implementation on the test vehicle are discussed.

\vspace{-4pt}
\subsection{Test Vehicle Instrumentation}\vspace{-4pt}
The eco-driving and eco-heating experiments are performed on a 2017 Toyota Prius Four Turing HEV. The test vehicle has been instrumented to (i) collect powertrain and HVAC data, and (ii) control the inputs (blower flow rate and cabin temperature setpoint) to the HVAC system using an external processor/laptop. Additional thermocouples have been installed to measure $T_{ain}$ and cabin temperatures ($T_{cab}$) at different locations, e.g., dashboard, roof, and glass. In order to override the HVAC system inputs by an external processor, the original HVAC control loops are modified to access the available signals on the controller area network (CAN) and local interconnect network (LIN) buses through an In-Car PC and a NeoVI Fire module. The external processor/laptop communicates with the In-Car PC via an Ethernet adaptor. The  schematic  of  the  modified  Prius HVAC system is shown in Figure~\ref{fig:PriusInterface}. 
\begin{figure}[t!]
\begin{center}
\includegraphics[angle=0,width=1\columnwidth]{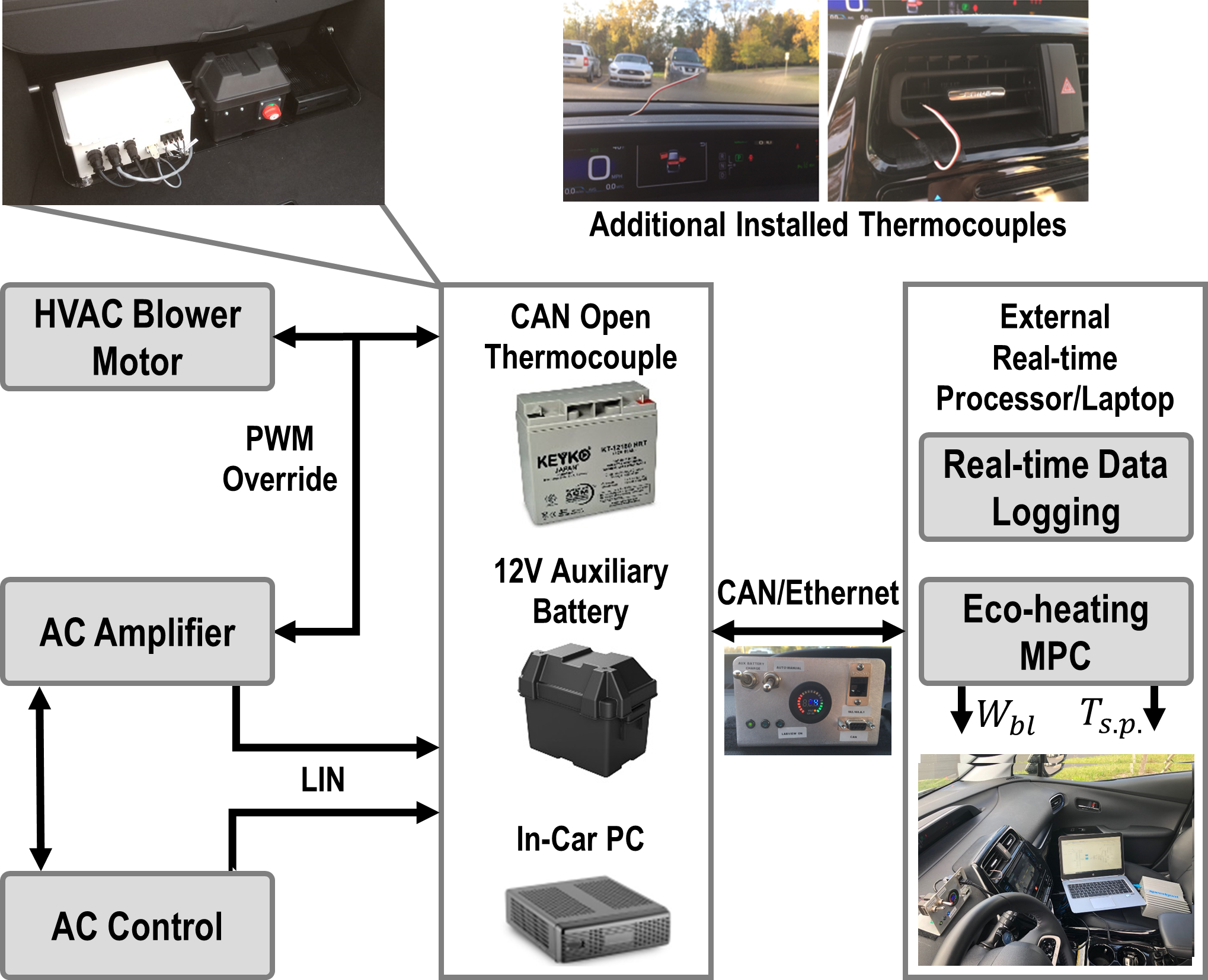} \vspace{-0.55cm}
\textcolor{Blue}{\caption{\label{fig:PriusInterface} Schematics of the modified Toyota Prius HVAC control system with
added thermocouples, CAN open thermocouple module, auxiliary battery, and In-Car PC.}}\vspace{4pt}
\end{center}
\end{figure}

\vspace{-4pt}
Note that the actual control input to the HVAC system for the blower flow rate ($W_{bl}$) is the percentage pulse width modulation (PWM) signal. This is while $\dot{m}_{bl}$ was used in the formulation of the eco-heating MPC in Eqs.~(\ref{eq:Eco-cooling}) and (\ref{eq:DAHP}). The blower airflow map is calibrated and the results are shown in Figure~\ref{fig:fig_blwer_flow}. This map incorporated as a look-up table in the eco-heating strategy to convert the output of MPC (\ref{eq:Eco-cooling}) to $W_{bl}$ percentage. 
\begin{figure}[h!]
\begin{center}
\includegraphics[width=1\columnwidth]{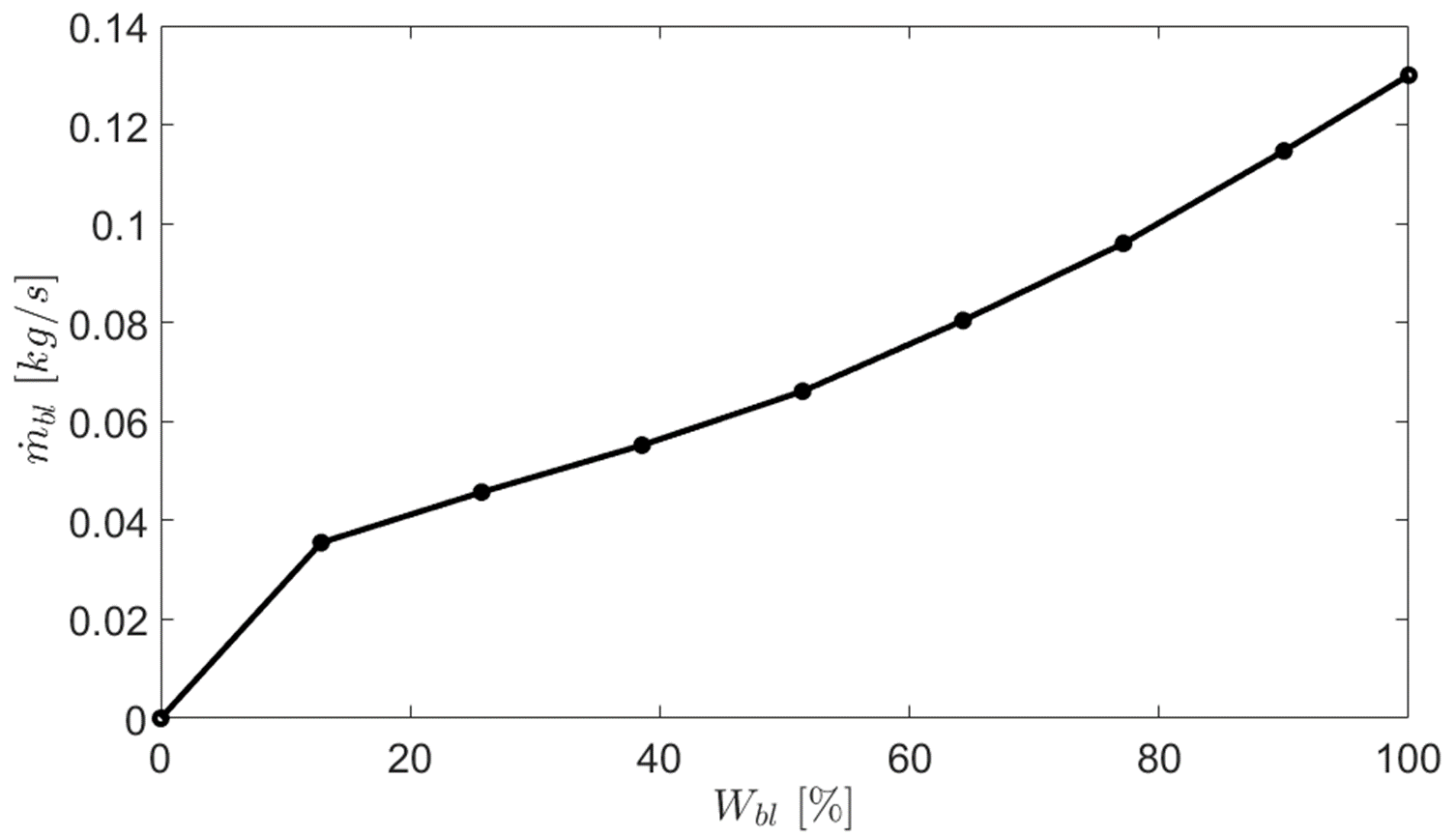}  
\end{center}\vspace{-0.5cm}
\textcolor{Blue}{\caption{\label{fig:fig_blwer_flow} HVAC blower airflow map.}}\vspace{-10pt}
\end{figure} 

\vspace{-4pt}
A Simulink/Stateflow\textsuperscript{\textregistered} interface was also developed to facilitate the design and testing of the eco-driving and eco-heating strategies. This interface parses the CAN messages received from the vehicle, including the feedback signals from the HVAC system, and converts the control signals and sends them to the HVAC system. 

\vspace{-4pt}
The Prius is driven by a human driver. To mimic a CAV driving scenario,~the developed Simulink interface provides a preview of the future planned speed profile on a screen to guide the driver to follow the speed profile. This speed advisory interface is shown in Figure~\ref{fig:Test_interface}, where the current speed and future planned speed are displayed as a red dot and blue line on the screen, respectively. The vehicle speed preview is updated in a receding scheme towards the end of the test period.\vspace{-4pt}
\begin{figure}[h!]
\begin{center}
\includegraphics[angle=0,width=0.9\columnwidth]{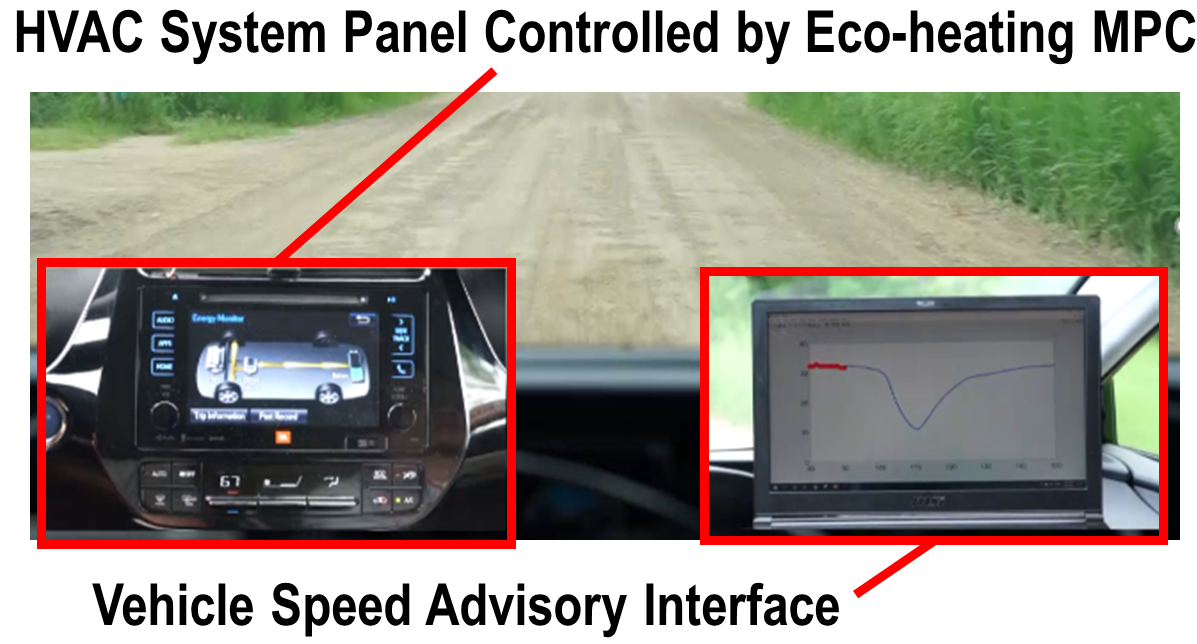} \vspace{-0.35cm}
\textcolor{Blue}{\caption{\label{fig:Test_interface} The vehicle speed advisory interface designed for better speed following by the human driver.}}\vspace{-5pt}
\end{center}
\end{figure}

\vspace{-4pt}
Finally, the vent air temperature ($T_{ain}$) in Eq.~(\ref{eq:DAHP}) is estimated based on the coolant temperature ($T_{cl}$) and cabin air temperature setpoint ($T_{s.p.}$) as follows:\vspace{-4pt}
\begin{eqnarray}
\label{eq:T_ain}
T_{ain}=\alpha_1 T_{s.p.}+\alpha_2(T_{sp}-\underline{T_{s.p.}})T_{cl}+\alpha_3 T_{s.p.}^2+\alpha_4
\end{eqnarray} 
where $\underline{T_{s.p.}}=15^oC$ is the minimum allowable cabin temperature setpoint on the Prius HVAC system. $\alpha_{1,2,3,4}$ are constant coefficients that are identified based on the data collected from the Prius. Figure~\ref{fig:TainValidation} shows validation of $T_{ain}$ model in Eq.~(\ref{eq:T_ain}) against data collected from the Prius with reasonable accuracy given the complex dynamics involved.
\begin{figure}[h!]
\begin{center}
\includegraphics[angle=0,width=1\columnwidth]{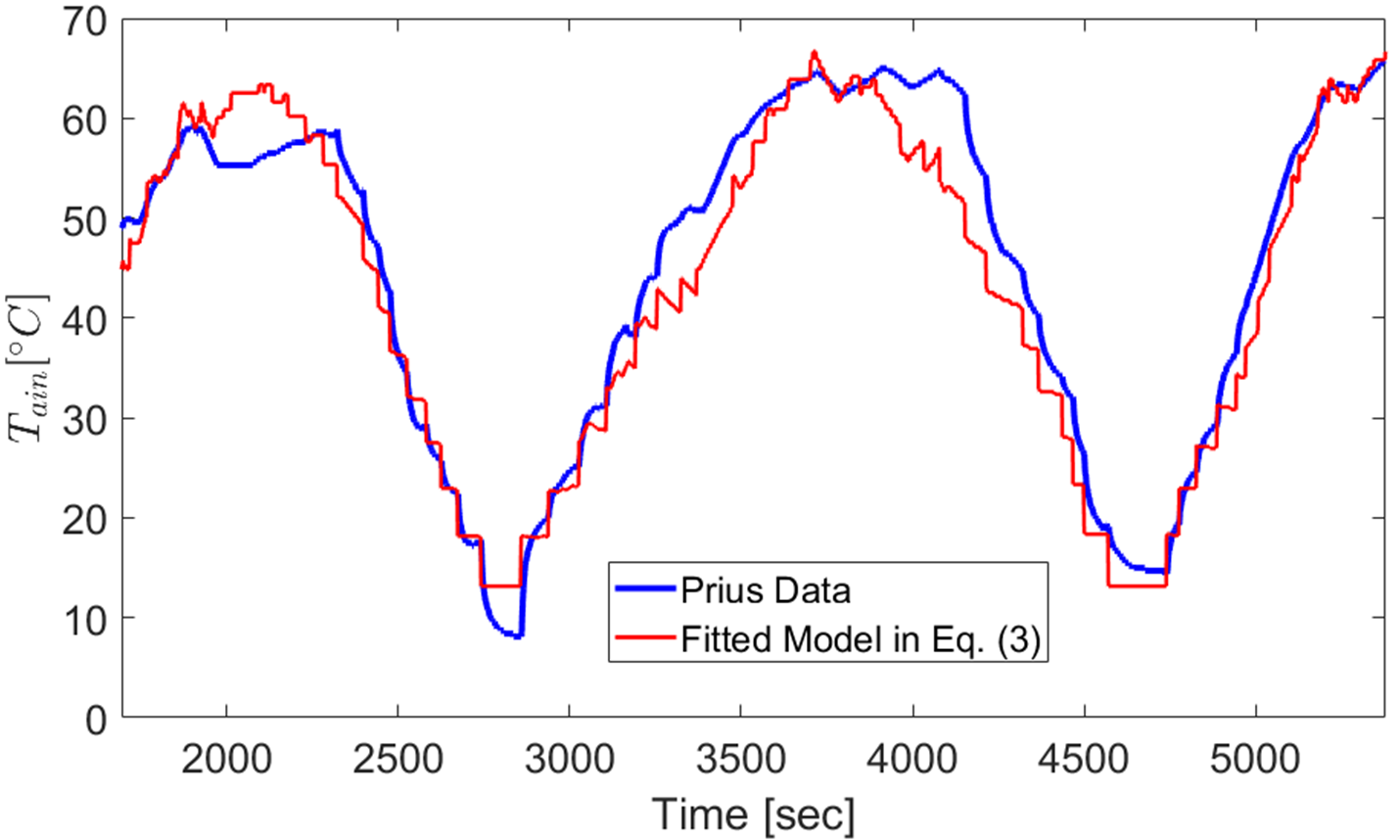} \vspace{-0.75cm}
\textcolor{Blue}{\caption{\label{fig:TainValidation} Model validation results of $T_{ain}$ for the sinusoidal excitation applied to the HVAC system on the test vehicle.}}\vspace{-4pt}
\end{center}
\end{figure}

\vspace{-4pt}
\subsection{Procedure of Eco-driving and Eco-heating Experiments}\vspace{-4pt}
A pair of vehicle speed profiles over Plymouth Rd. are selected for vehicle testing. The testing has been performed in two phases. During the first phase, the focus was to demonstrate eco-driving benefits by comparing the Prius energy consumption with normal-driving and with eco-driving. The second testing phase was focused on eco-heating, for which the eco-trajectory was used as the speed profile to follow.~For safety and repeatability, a road outside of Ann Arbor with similar length to that of Plymouth Rd. corridor and with light traffic was selected to perform the tests. The testing road, South Zeep Rd, and the actual Plymouth Rd. are shown in Figure~\ref{fig:Plymouth_Zeeb}. Virtual signalized intersections were considered along the testing road to emulate the six intercessions along the Plymouth Rd.
\begin{figure}[h!]
\begin{center}
\includegraphics[angle=0,width=1\columnwidth]{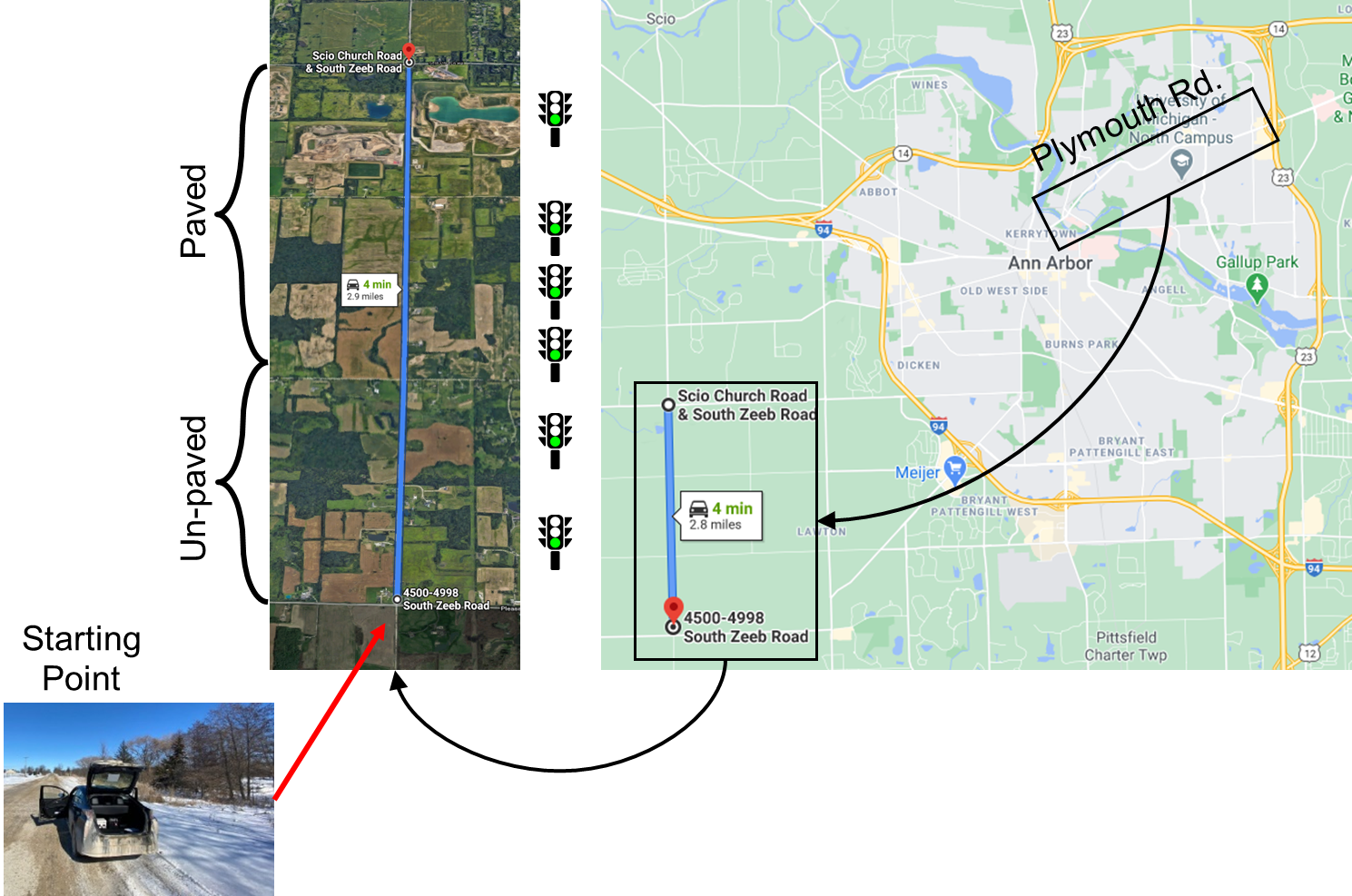} \vspace{-0.75cm}
\textcolor{Blue}{\caption{\label{fig:Plymouth_Zeeb} The South Zeeb Rd. selected for testing based on the speed profiles generated for Plymouth Rd. with virtual signalized intersections.}}\vspace{-4pt}
\end{center}
\end{figure}

\vspace{-4pt}
\subsubsection{Phase I: Eco-driving Experiments} 
\vspace{-4pt}
The speed profiles selected for eco-driving experiments are shown in Figure~\ref{fig:PlymouthSpeed}. The eco-trajectory is shown in Figure~\ref{fig:PlymouthSpeed}-(a). Figure~\ref{fig:PlymouthSpeed}-(b) is for the same vehicle traveling normally by a human driver, i.e., its speed is not optimized using the eco-trajectory planning algorithm. As an example of the difference between normal-driving and eco-driving, one can notice that upon departure from the third intersection around $t=220~sec$, the normal-driving vehicle accelerates to approach the fourth intersection. Due to the left-over queue at the fourth intersection, the normal-driving vehicle has to decelerate rapidly around $t=250~sec$, then it starts accelerating again once the queue is discharged. The eco-driving vehicle, on the other hand, avoids such aggressive acceleration/deceleration behavior by leveraging the V2I information that informs it about the queue condition at the fourth intersection. {Note that, for the normal driving scenarios, an incorporated human driver model in VISSIM has been used to generate the speed profile. This model tries to mimic human driver behavior in tracking a given cruise speed while assuming that the human driver cannot follow the reference speed exactly, resulting in small fluctuations in the speed profile---as can be seen in Figure~\ref{fig:PlymouthSpeed}-(b).} 
\begin{figure}[h!]
\begin{center}
\includegraphics[angle=0,width=1\columnwidth]{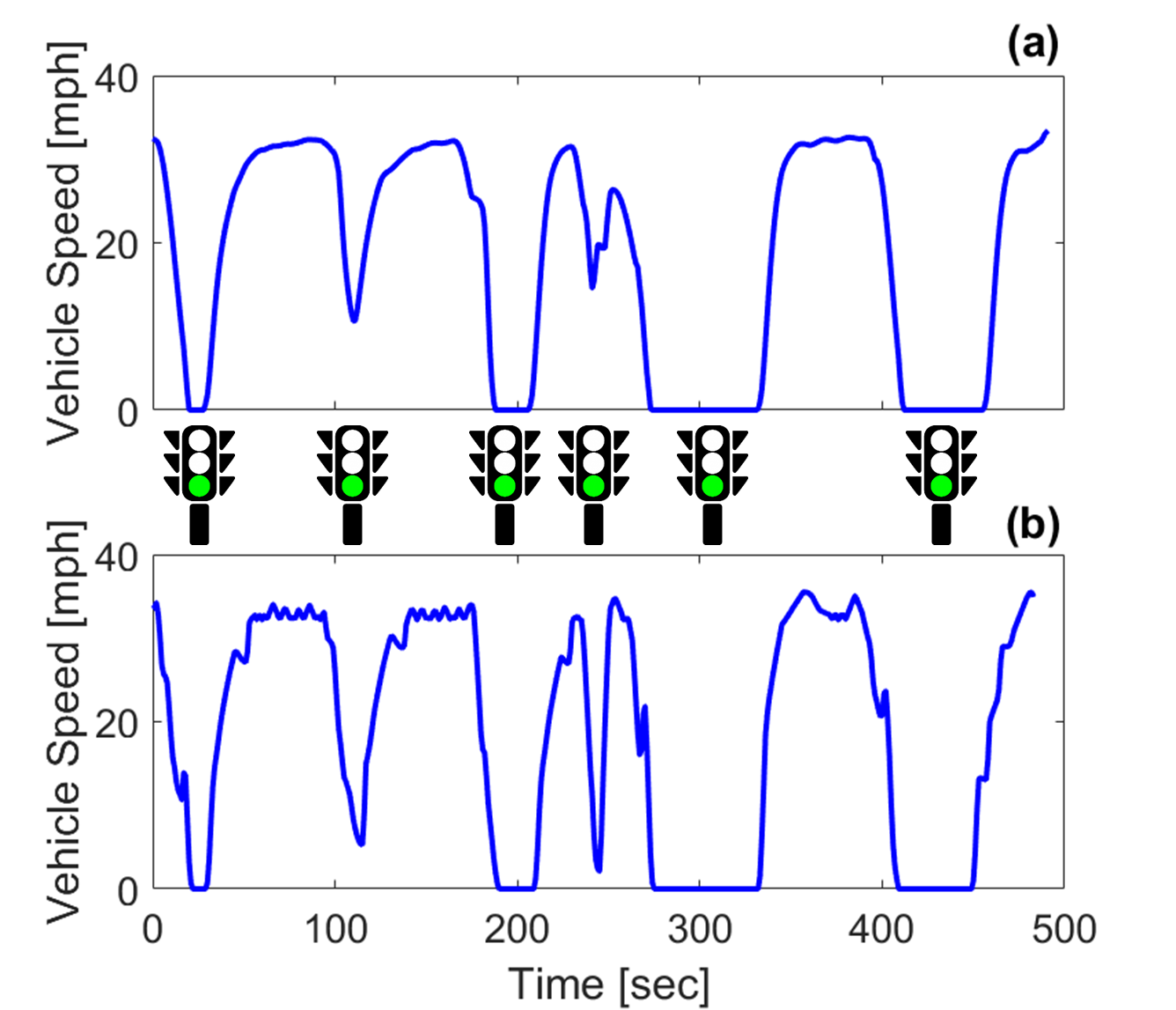} \vspace{-0.85cm}
\textcolor{Blue}{\caption{\label{fig:PlymouthSpeed} Speed profiles used for eco-driving and eco-heating experiments: (a) eco-trajectory with EAD, and (b) normal driving with human driver.}}\vspace{-15pt}
\end{center}
\end{figure}

\begin{table*}
\textcolor{Blue}{\caption{\label{tab:my_label} The summary of speed tracking performance.}}
\centering
    \begin{tabular}{|c|c|c|}
    \hline
    \textbf{Normal-driving}   & \textbf{Average speed tracking}  & \textbf{Standard deviation} \\
    & \textbf{absolute error} [$mph$] & \textbf{of the speed tracking error} [$mph$] \\ \Xhline{5\arrayrulewidth}
    Test 1 & 1.9 & 3.3 \\ \hline
    Test 2 & 1.7 & 2.8 \\ \hline
    Test 3 & 1.5 & 2.4 \\ \hline
    Test 4 & 1.5 & 2.5 \\ \Xhline{5\arrayrulewidth}
    \textbf{Eco-driving}   &  & \\ \hline 
    Test 1 & 1.4 & 2.2 \\ \hline
    Test 2 & 1.2 & 1.9 \\ \hline
    Test 3 & 1.2 & 2.0 \\ \hline
    Test 4 & 1.1 & 1.9 \\ \hline 
    \end{tabular}\vspace{-7pt}
\end{table*}

\vspace{-4pt}
Overall, 20 vehicle tests (some test data had to be disregarded due to the disruption by unexpected traffic or other reasons) were conducted; 10 tests with eco-driving and 10 tests with normal-driving. The tests were performed during January and February of 2019, with the first test completed on January 6th, and the last one on February 3rd. The vehicle HVAC system was set to provide constant heating to the cabin, minimizing the impact of HVAC operation on the eco-driving results. The driver performance in tracking the speed profile is listed in Table~\ref{tab:my_label} for 8 tests (4 eco-driving, 4 normal driving). With the average error being bounded by 2 $mph$, the tests showed acceptable speed tracking performance. Note that the standard deviation for normal driving is slightly higher ($<1~mph$). This is mainly because the speed profile for normal-driving (Figure~\ref{fig:PlymouthSpeed}-(b)) is less smooth, making it hard for the driver to follow the target speed.

\vspace{-6pt}
\subsubsection{Phase II: Eco-heating Experiments} 
\vspace{-4pt}
For the eco-heating experiments, the speed profile is fixed to be the eco-trajectory shown in Figure~\ref{fig:PlymouthSpeed}-(a). The HVAC system is tested in two scenarios as shown in Figure~\ref{fig:TestScenario}. In the first scenario, the inputs of the HVAC system are fixed at $T_{s.p.}=23^oC$ and $W_{bl}=40\%$. In the second scenario, the MPC in Eq. (\ref{eq:Eco-cooling}) is used to compute and apply the control inputs according to the speed preview, which is assumed to be a known a priori over the receding prediction horizon of $N_p=30~sec$. {The ranges for the control inputs $T_{s.p.}$ and $W_{bl}$ are defined as $18^oC\leq T_{s.p.}\leq 24^oC$ and $10\%\leq W_{bl}\leq 70\%$.} The MPC is solved with a sampling time of $\delta t=1~sec$.

\vspace{-4pt}
It should be noted that energy saving is not expected from the heating system itself. Rather it is achieved by coordinating the heating and powertrain systems. Specifically, energy can be saved
by reducing the engine idling time. A total number of 16 tests (more tests were conducted, but part of the collected data had to be disregarded to ensure consistency and comparability in the reported results) were conducted over the same testing road shown in Figure~\ref{fig:Plymouth_Zeeb}. The testing period was from December 2019 to February 2020. During this period, weather conditions were varying, with ambient temperature ranging from -11$^oC$ to +7$^oC$. Each constant-heating experiment was followed by an eco-heating experiment to make the data sets comparable with similar ambient temperatures. Each test started at the same location, with an initial cabin temperature between 13 to 15 $^oC$, and a fully warmed up engine.

\vspace{-4pt}
\section{Experimental Results and Discussions}\vspace{-4pt}
Figure~\ref{fig:Eco-driving_results} presents the average energy consumption of all the eco-driving tests performed on the Prius. To adjust the fuel consumption to account for the difference between the final battery state-of-charge ($SOC(t_f)$) and its initial value ($SOC(0)$), the equivalent total energy consumption ($E_{eq}$) is calculated using:\vspace{-2pt}
\begin{gather}
\label{eq:Eveh}
E_{eq} = \int_{0}^{t_{f}}\frac{\dot{m}_{air}(t)}{AFR(t)}\cdot LHV~dt + \frac{E_{batt} \cdot \Delta SOC}{\eta_{sys}},
\end{gather}  
where $\dot{m}_{air}$ is the air flow rate into the engine measured by the mass air flow (MAF) sensor, $AFR$ is the air-fuel ratio, $LHV$ represents the lower heating value (LHV) of the gasoline, $E_{batt}$ is the battery capacity, $\eta_{sys}$ is the energy conversion efficiency from fuel energy to battery energy, and:\vspace{-2pt}
\begin{gather}
\label{eq:Eveh_2}
AFR(t) =  \lambda(t)\cdot AFR_{stoich},\\
\Delta SOC = SOC (0)- SOC (t_f),
\end{gather}
in which $\lambda$ and $AFR_{stoich}$ are the equivalent air-fuel ratio (measured), and the stoichiometric air-fuel ratio, respectively. \vspace{-6pt}
\begin{figure}[h!]
\begin{center}
\includegraphics[angle=0,width=0.8\columnwidth]{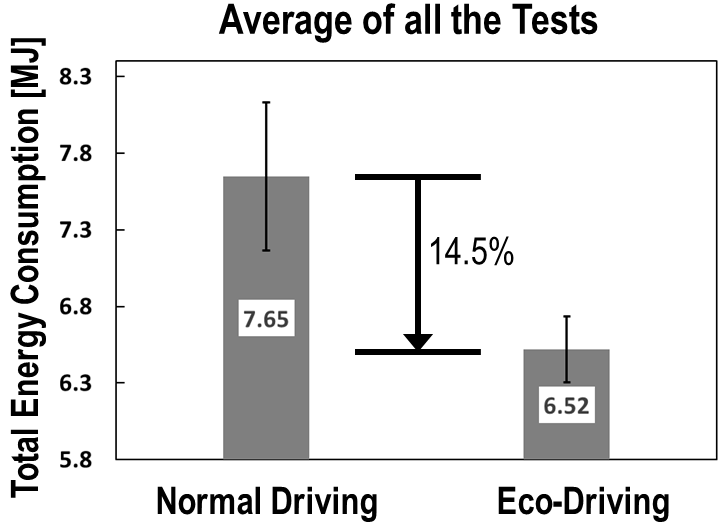} \vspace{-0.35cm}
\textcolor{Blue}{\caption{\label{fig:Eco-driving_results} The summary of eco-driving test results and its impact on the Prius energy consumption. The bars show the standard deviations.}}
\end{center}
\end{figure}

\begin{figure*}
\begin{center}
\includegraphics[angle=0,width=1.45\columnwidth]{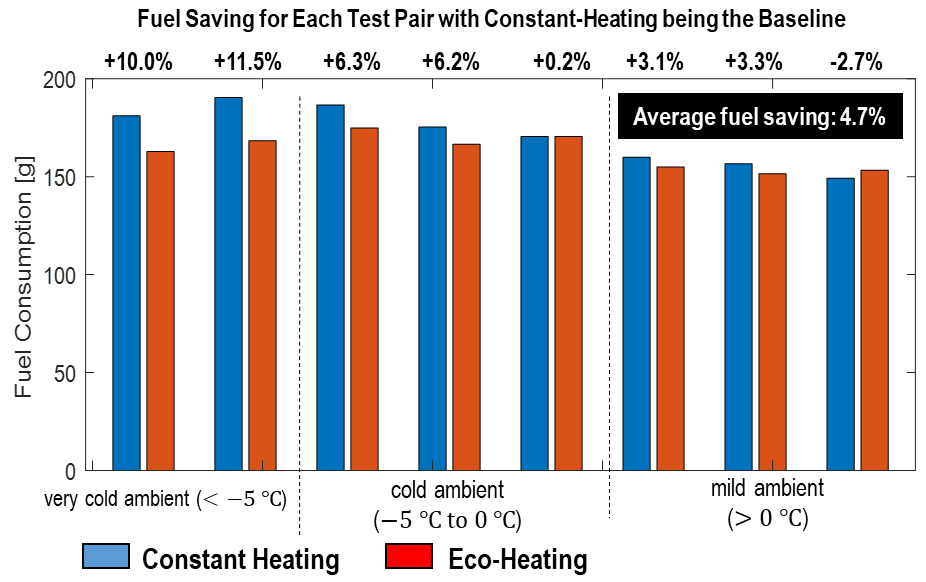} \vspace{-0.35cm}
\textcolor{Blue}{\caption{\label{fig:Eco-heating_results} The summary of eco-heating test results and its impact on the Prius energy consumption. In all cases, the eco-trajectory generated for eco-driving over the Plymouth Rd. corridor was used.}}
\end{center}
\end{figure*}

\vspace{-4pt}
For the particular speed profile selected for eco-driving testing, our previous simulation-based study in~\cite{AminiACC19} showed an energy saving of 18$\%$ achieved through eco-driving. The test results in Figure~\ref{fig:Eco-driving_results} show an average saving of 14.5$\%$ resulted from 20 experiments. The difference between the simulation and tests can be explained by considering (i) the error in tracking the target speed by the human driver, (ii) partially unpaved testing road with higher traction losses, and (iii) the differences between the Prius power-split controller as compared to the one used in the vehicle simulation model. While in the simulation environment the power-split logic is known and can be adjusted, such strategy on the test vehicle is not known as its details are proprietary to original equipment manufacturers (OEMs).

\vspace{-4pt}
The (adjusted) fuel consumption results of eco-heating experiments with different ambient temperatures are listed in Figure~\ref{fig:Eco-heating_results}, and are compared with constant heating cases at the same ambient temperature.~The ambient temperatures are categorized into three groups, namely very cold ambient ($T_{amb}\leq-5^oC$), cold ambient ($-5<T_{amb}\leq0~^oC$), and mild ambient ($T_{amb}>0^oC$). While the eco-heating strategy shows an average fuel saving of 4.7\% for these group of experiments, Figure~\ref{fig:Eco-heating_results} shows that he benefits of eco-heating are more pronounced for colder ambient temperatures.

\begin{figure}[t!]
\begin{center}
\includegraphics[angle=0,width=1\columnwidth]{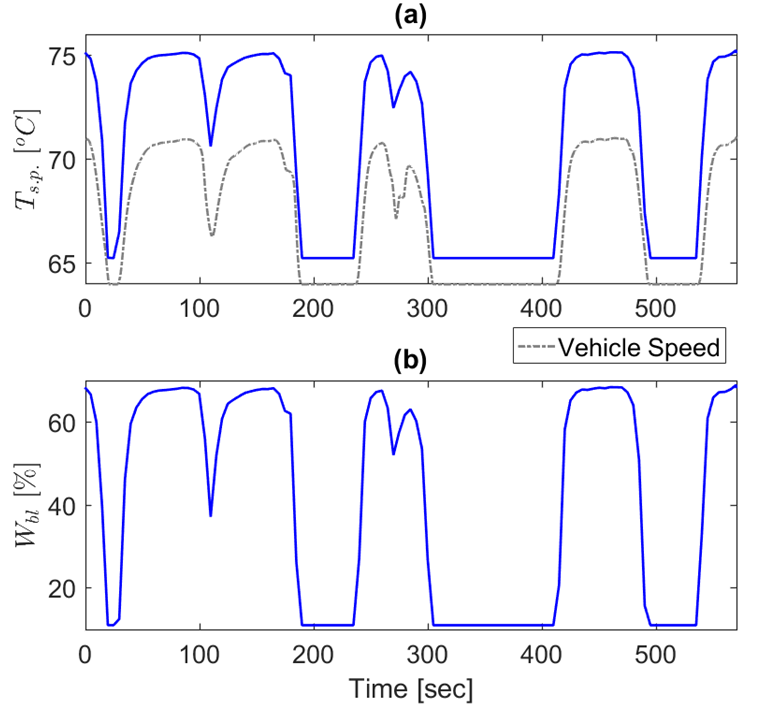} \vspace{-0.75cm}
\textcolor{Blue}{\caption{\label{fig:eco_heat_inputs} The computed HVAC control inputs by the MPC of the eco-heating strategy: (a) $T_{s.p.}$, and (b) $W_{bl}$.}}\vspace{-10pt}
\end{center}
\end{figure}

\begin{figure}[t!]
\begin{center}
\includegraphics[angle=0,width=1\columnwidth]{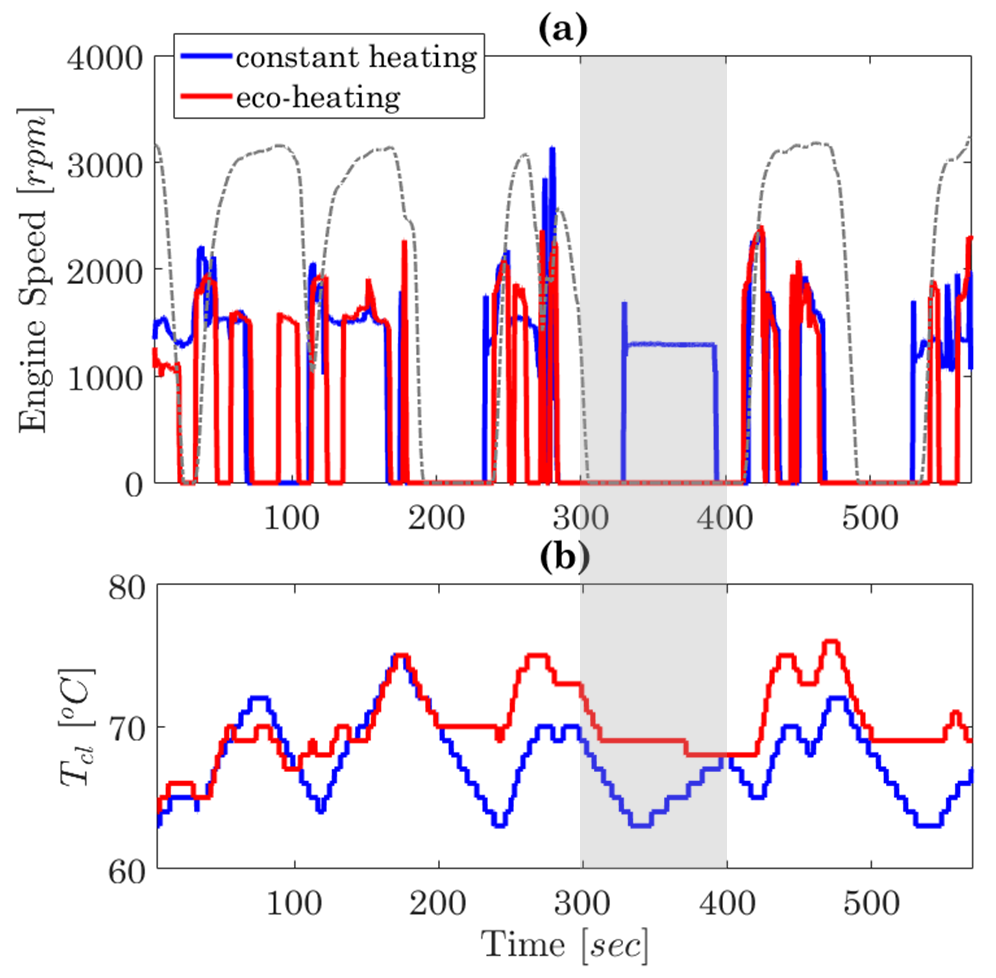} \vspace{-0.95cm}
\textcolor{Blue}{\caption{\label{fig:Eco_heating_sample} (a) Engine speed and (b) coolant temperature trajectories with and without eco-heating strategy for two back-to-back experiments at the same ambient temperature of $T_{amb}=-8^oC$.}}\vspace{-6pt}
\end{center}
\end{figure}

\begin{figure}[h!]
\begin{center}
\includegraphics[angle=0,width=1\columnwidth]{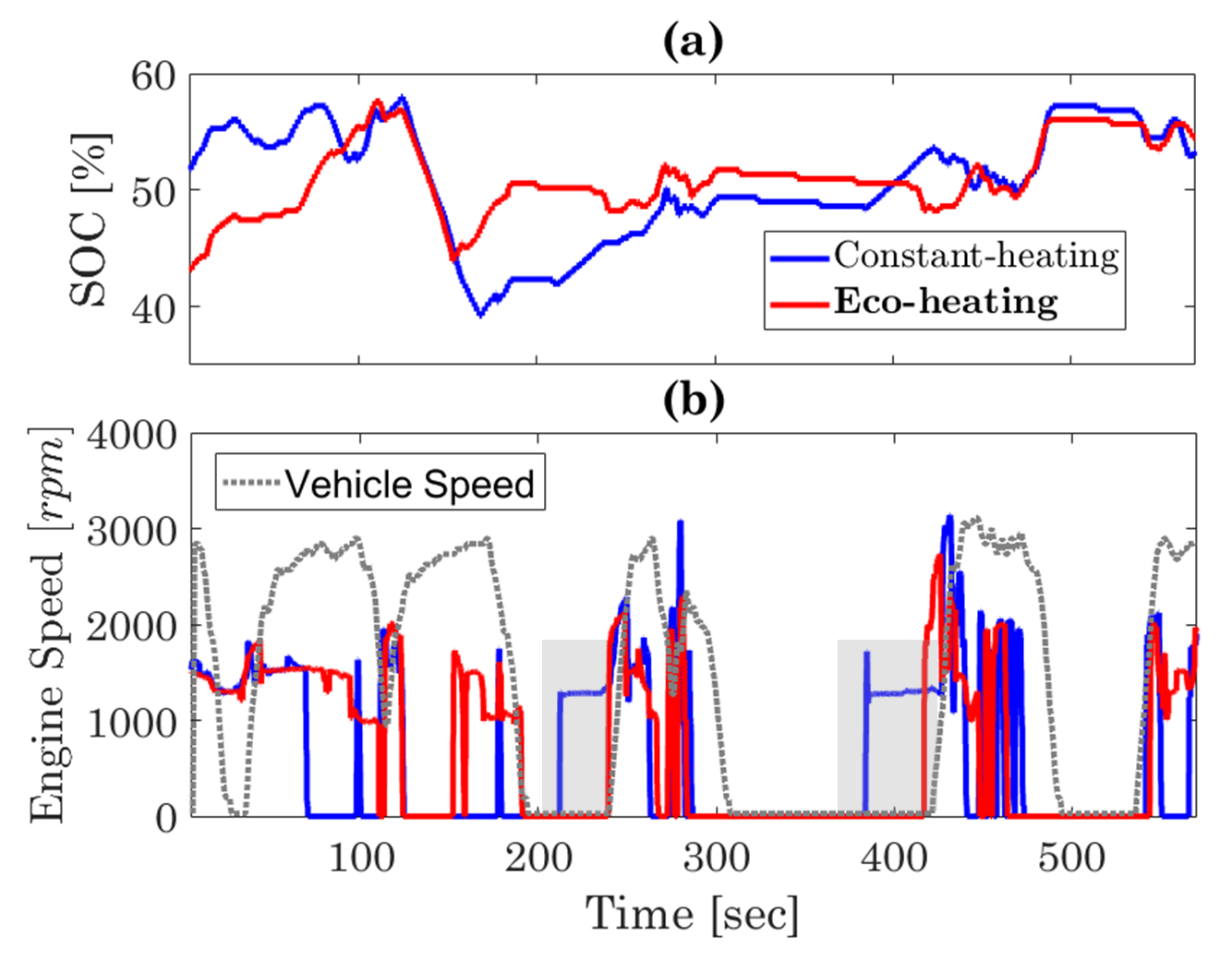} \vspace{-0.95cm}
\textcolor{Blue}{\caption{\label{fig:Eco_heating_sample2} (a) Battery $SOC$ and (b) engine speed trajectories with and without eco-heating strategy for two back-to-back experiments at the same ambient temperature of $T_{amb}=-11^oC$. The highlighted box in subplot (b) shows the engine idling period.}}\vspace{-4pt}
\end{center}
\end{figure}

\vspace{-4pt}
The computed HVAC control inputs by the MPC (Eq. ~(\ref{eq:Eco-cooling})) of the eco-heating strategy is shown in Figure~\ref{fig:eco_heat_inputs}. The speed profile used for testing is also shown in Figure~\ref{fig:eco_heat_inputs}-(a), which is the same as the eco-trajectory used for eco-driving experiments with the third stop being extended to better highlight the impact of cabin heating on the engine idling. To explain the fuel-saving achieved through eco-heating, the engine speed and coolant temperature trajectories recorded from one pair of sequential experiments are plotted in Figure~\ref{fig:Eco_heating_sample}. During the long vehicle stop between $t=$300 to 400 $sec$, the coolant temperature with constant-heating drops quickly due to the heating demand. At around $T_{cl}=$63$^oC$, the vehicle controller commands the engine to be turned on to provide heat. Eco-heating strategy, on the other hand, foresees such an inefficient operation according to the incorporated look-ahead information. It uses the coolant as the thermal storage to store the thermal energy before the vehicle's stop by increasing $W_{bl}$ and $T_{s.p.}$ as shown in Figure~\ref{fig:eco_heat_inputs}. This results in a higher coolant temperature at $t=300~sec$, giving enough thermal energy in the coolant to heat the cabin without forcing the engine to idle. At the same time, to ensure the coolant thermal energy is not being depleted quickly, the MPC reduced the heat transfer rate from the coolant to the cabin by decreasing $W_{bl}$ and $T_{s.p.}$.  

\vspace{-4pt}
Here, we report two additional special cases to further highlight the performance of the eco-heating strategy, one at very cold ambient and the other one at relatively mild ambient. Figure~\ref{fig:Eco_heating_sample2} shows the powertrain trajectories for the pair of experiments at $T_{amb}=-11^oC$. At this cold ambient temperature, it can be seen that engine idling occurs during two stops with constant-heating HVAC setting around $t=250~sec$ and $410~sec$. Eco-heating strategy avoids such inefficient use of engine during both of these vehicle stops. Figure~\ref{fig:Eco_heating_sample2}-(a) shows that during the engine idling period, as expected, the battery is also being charged. For this pair of experiments, the eco-heating strategy reduced the adjusted fuel consumption by 2.5\%, as compared to constant-heating case. This saving is archived while $E_{DAHE}$ is increased by 2.2\% from 1.86 $MJ$ (constant-heating) to 1.9 $MJ$ (eco-heating). The latter observation shows that fuel saving can be achieved without sacrificing (even increasing) the total heating delivered to the cabin, confirming the benefits of eco-heating.

\vspace{-4pt}
Finally, we report a special case at relatively mild ambient temperature of $T_{amb}=+6~^oC$ in Figure~\ref{fig:Eco_heating_sample3}. Figure~\ref{fig:Eco_heating_sample3}-(c) shows that the eco-heating strategy maintains the coolant temperature at a constant level during the vehicle stops. With the constant-heating strategy, the coolant temperature drops during the vehicle stops. Since the ambient temperature is mild, engine idling was not observed in either of these cases as shown in Figure~\ref{fig:Eco_heating_sample3}-(b). With different initial cabin temperatures (the difference is less than 3$^oC$), as shown in Figure~\ref{fig:Eco_heating_sample3}-(d), a similar cabin temperature is observed from both strategies after $t=100~sec$ despite having different coolant temperatures. The eco-heating strategy reduced the adjusted fuel consumption by 2.8\%. $E_{DAHE}$, however, decreased by 38\% from 1.69 $MJ$ (constant-heating) to 1.04 $MJ$ (eco-heating). The latter observation suggests that the eco-heating strategy may not provide much saving at mild ambient temperature as it could result in a reduced $E_{DAHE}$. This special case shows further investigation is required for a better understanding of passengers comforts and its relationship with fuel economy to address the trade-off between comfort and fuel economy.~It should be noted that the presented results here are based on comparisons against a constant-heating scenario with fixed HVAC inputs. The comparison against the ``AUTO'' mode developed by the OEM may lead to different numbers.

\vspace{-4pt}
\section{Summary and Conclusions}\label{Stage_II_Conclusion}\vspace{-4pt}
In this paper, we presented the experimental validation results of a combined eco-driving and eco-heating strategy developed for CAVs operating in congested urban roads. The eco-driving strategy aims at optimizing the vehicle speed based on (i) V2I communication, and (ii) estimation of the queuing dynamics at signalized intersections. The optimized speed profile by the eco-driving strategy was then used as look-ahead information to develop the eco-heating strategy, which is based on model predictive control. The eco-heating strategy aims at coordinating the cabin heating load with vehicle speed. Such coordination allows for leveraging the engine coolant and cabin air as thermal energy storage. By shifting the cabin thermal loads away from the periods where the engine operation is not efficient, the eco-heating strategy minimizes the need for running the engine during low vehicle speeds and saves fuel.
\begin{figure}[t!]
\begin{center}
\includegraphics[angle=0,width=1\columnwidth]{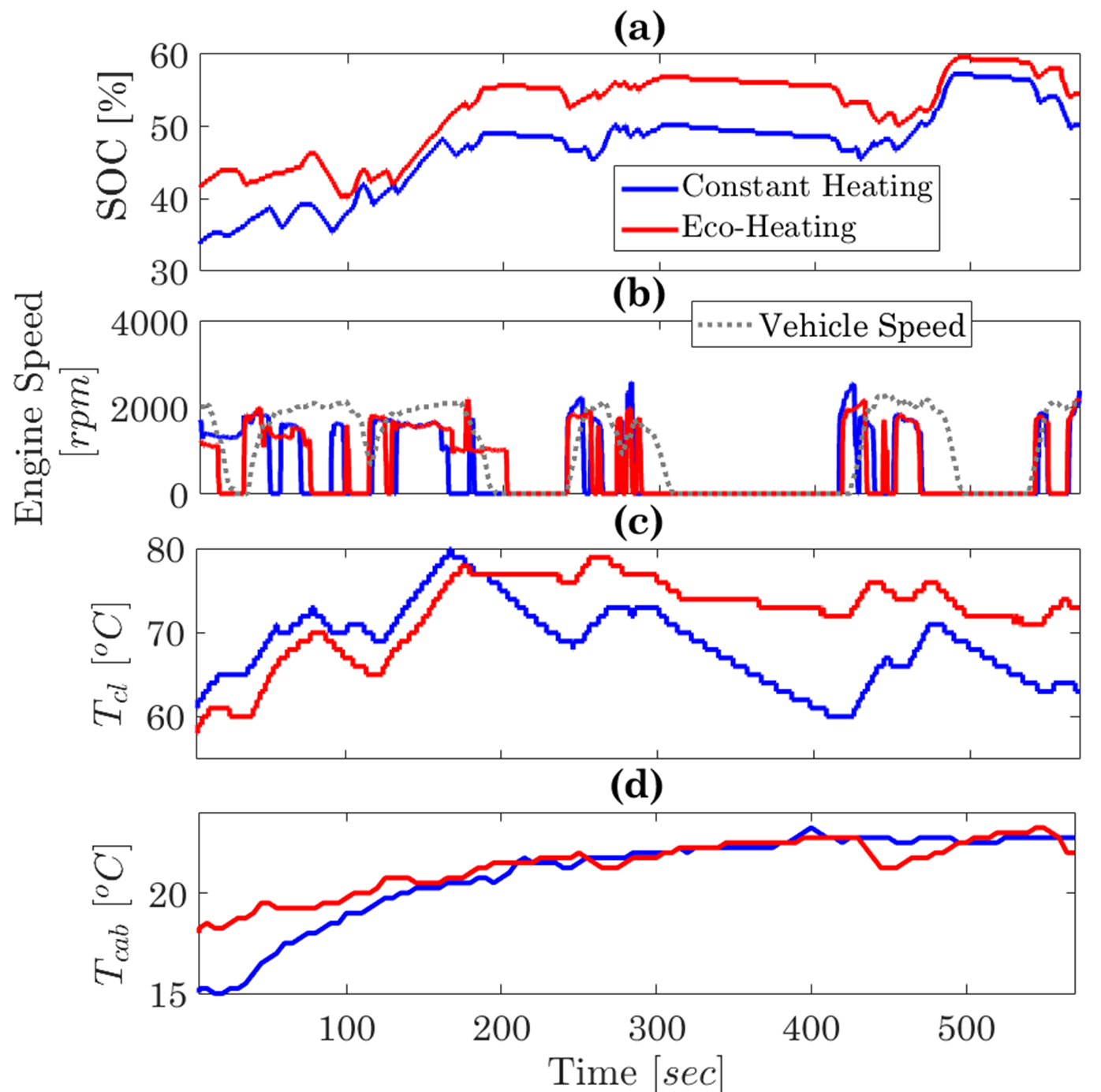} \vspace{-0.75cm}
\textcolor{Blue}{\caption{\label{fig:Eco_heating_sample3} (a) Battery $SOC$, (b) engine speed, (c) coolant temperature $T_{cl}$, and (d) cabin temperature $T_{cab}$ trajectories with and without eco-heating strategy for two back-to-back experiments at the same ambient temperature of $T_{amb}=+6^oC$.}}\vspace{-6pt}
\end{center}
\end{figure}

\vspace{-4pt}
Extensive experiments were performed on a 2017 Toyota Prius Four Turing HEV. The test vehicle allows for real-time powertrain data logging and HVAC inputs overriding from an external computer. The eco-driving experiments showed an average fuel saving of 14.5\% for driving on an arterial corridor in Ann Arbor, MI. Eco-heating experiments over the same corridor showed an additional saving of 4.5\% on average, offering a combined saving of up to 19\% achieved through eco-driving and eco-heating when compared with a baseline case with normal driving by a human driver and with constant heating. The fuel-saving was achieved by the eco-heating strategy with consideration of the passenger comfort and total heating energy delivered to the cabin. Future works will focus on developing a more accurate model representing passenger comforts in the eco-heating strategy. Performing the experiments in a controlled environment will reduce the sources of uncertainties in the results, particularly for the eco-heating strategy which depends on the ambient temperature.~{Moreover, in this paper, only the baseline Prius HVAC controller with constant settings was considered. More comprehensive work is needed to fully quantify the performance of the benchmark HVAC controller and identify the opportunities for energy-saving through connectivity-enabled technologies.}~These topics are left for future investigations. 

\vspace{-4pt}
\section{Contact Information}\vspace{-4pt}
{Mohammad Reza Amini}, \underline{mamini@umich.edu}

\vspace{-4pt}
{Hao Wang\footnote{H. Wang's contribution occurred while at the University of Michigan.}}, \underline{hwang210@ford.com}

\vspace{-4pt}
{Ilya Kolmanovsky}, \underline{ilya@umich.edu} 

\vspace{-4pt}
{Jing Sun}, \underline{jingsun@umich.edu} 

%
\vspace{-4pt}
\section*{Acknowledgments}\vspace{-4pt}
The University of Michigan portion of this work is funded in part by the United States Department of Energy (DOE), ARPA-E NEXTCAR program under award No. DE-AR0000797.\vspace{-4pt}

\bibliographystyle{unsrt}
\bibliography{SAE2020_bib}
\onecolumn

\section*{Appendix}
\vspace{-8pt}
\emph{\textbf{Nomenclature}}:\\
\vspace{-10pt}
\begin{table}[h]
\addvspace{0.2in}
\renewcommand\arraystretch{1.2}
\begin{tabular}{l l}
$AFR$ & Air-fuel ratio, [$-$]\\ 
$E_{DAHP}$ & Discharge air heating energy, [$J$]\\ 
$E_{eq}$ & Equivalent total energy consumption, [$J$]\\
$k$ & Discrete time index, [$-$]\\
$\dot{m}_{air}$ & Air flow rate into
the engine, [$kg/sec$]\\ 
$\dot{m}_{bl}$ & Blower air mass flow rate, [$kg/sec$]\\ 
$N_p$ & Prediction horizon, [$step$]\\ 
$P_{DAHP}$ & Discharge air heating power, [$W$]\\ 
$P_{DAHP}^{targ}$ & Target discharge air heating power, [$W$]\\ 
$SOC$ &Battery state-of-charge, [$\%$]\\ 
$T_{ain}$ &Vent air temperature, [$^oC$]\\
$T_{amb}$ &Ambient temperature, [$^oC$]\\
$T_{cl}$ &Engine coolant temperature, [$^oC$]\\
$T_{s.p.}$ &Cabin temperature setpoint, [$^oC$]\\
$t$ & Time, [$^secC$]\\
$t_f$ & Final time, [$sec$]\\
$v$ &Vehicle speed, [$m/sec$]\\ 
$W_{bl}$ & Blower air flow rate percentage, [$\%$]\\ 
$\delta t$ &Sampling time, [$sec$]\\
\vspace{-0.16cm}
\end{tabular}
\end{table}

\vspace{-15pt}
\emph{\textbf{Acronyms}}:\\
\vspace{-10pt}
\begin{table}[h]
\addvspace{0.2in}
\renewcommand\arraystretch{1.2}
\begin{tabular}{l l}
\textit{CAV} & Connected and automated vehicle \\
\textit{EAD} & Eco-arrival and eco-departure \\
\textit{EV} & Electric vehicle\\
\textit{HEV} & Hybrid electric vehicle\\
\textit{HECAV} & Hybrid electric connected and automated vehicle\\
\textit{HVAC} & Heating, ventilation and air conditioning\\
\textit{iPTM} &Integrated power and thermal management\\ 
\textit{LHV} & Lower heating value\\
\textit{MPC} & Model predictive control \\
\textit{V2I} & Vehicle to infrastructure \\
\textit{V2V} & Vehicle to vehicle \\
\vspace{-0.16cm}
\end{tabular}
\end{table}

\end{document}